\documentclass[12pt]{article}
\newcommand{\beq}{\begin{eqnarray}}
\newcommand{\eeq}{\end{eqnarray}}
\newcommand{\be}{\begin{equation}}
\newcommand{\ee}{\end{equation}}
\def\la{\mathrel{\mathpalette\fun <}}

\def\fun#1#2{\lower3.6pt\vbox{\baselineskip0pt\lineskip.9pt
\ialign{$\mathsurround=0pt#1\hfil ##\hfil$\crcr#2\crcr\sim\crcr}}}

\newcommand{{\SD}}{\rm SD}

\newcommand{\ver}{\mbox{\boldmath${\rm r}$}}

\newcommand{\vep}{\mbox{\boldmath${\rm p}$}}

\newcommand{\ves}{\mbox{\boldmath${\rm s}$}}

\newcommand{\veta}{\mbox{\boldmath${\rm \eta}$}}

\newcommand{\lan}{\langle}
\newcommand{\ran}{\rangle}
\begin{document}
\begin{center}

{\Large \bf Masses of the $\mathbf{\veta_c(nS)}$ and $\mathbf{\veta_b(nS)}$ mesons}
\vspace{1cm}

\large{ A.M.Badalian$^1$,  B.L.G.Bakker$^2$}\\

$^1$Institute of Theoretical and Experimental Physics, \\Moscow,
Russia\\
$^2$Department of Physics and Astronomy, Vrije Universiteit,
Amsterdam, The Netherlands
\end{center}
\vspace{1cm}

\begin{abstract}
\noindent
The hyperfine splittings in heavy quarkonia are studied using new
experimental data on the di-electron widths.  The smearing of the
spin-spin interaction is taken into account, while the radius of
smearing is fixed by the known $J/\psi-\eta_c(1S)$ and
$\psi(2S)-\eta'_c(2S)$ splittings and appears to be small, $r_{ss}
\approx 0.06$ fm.  Nevertheless, even with such a small radius an
essential suppression of the hyperfine splittings ($\sim 50\%)$ is
observed in bottomonium. For the $nS~ b\bar b$ states $(n=1,2,...6)$
the values we predict (in MeV) are 28, 12, 10, 6, 6, and 3,
respectively. In single-channel approximation for the $3S$ and $4S$
charmonium states the splittings 16(2) MeV and 12(4) MeV are
obtained.
\end{abstract}

\section{Introduction} 
At present two spin-singlet $S$-wave states $\eta_c (1S)$ and
$\eta_c(2S)$  are discovered  \cite{1}-\cite{3}; still, no spin-singlet
$\eta_b(nS)$ levels have been seen \cite{4}.  Though the masses of the
$\eta_b(nS)$ were predicted in many papers \cite{5}-\cite{11}, the
calculated hyperfine (HF) splittings, $ \Delta_{\rm HF} (nS) = M(n{}^3
S_1) - M(n{}^1S_0)$, vary in a wide range: from 35 MeV up to 100 MeV  for
the $b\bar b~ 1S$ state and for the $2S$ state between 19 MeV  and 44
MeV \cite{11}.  However, at the modern level of the theory and
experiment there exist well-established limits on the factors which
determine the spin-spin potential $V_{\rm HF}(r)$ in heavy quarkonia.
First of all, the wave function (w.f.) at    the origin for a given
$n~^3S_1(c\bar c $ or $b\bar b$) state can be extracted from
di-electron width which  are now measured with high accuracy
\cite{12,13}. Concerning the quark masses, the pole (current) mass,
present in a correct relativistic approach, and the constituent mass,
used in nonrelativistic or in more refined approximations, are also
known with good accuracy \cite{14,15}.  Therefore the only
uncertainties comes from two sources:

First, in perturbative QCD there is no  strict prescription how to
choose the renormalization scale $\mu$ in the strong coupling
$\alpha_{\rm HF}$, entering $V_{\rm HF}(r)$.

Secondly, the role of smearing of the spin-spin interaction is  not
fully understood and the true size of the smearing radius $r_{ss}$ is
still not fixed.

Moreover, the masses of higher triplet and singlet states can be
strongly affected by open channel(s), thus modifying the HF
splittings.

In our calculations the smearing radius $r_{ss}$ is taken to fit the
$J/\psi-\eta_c(1S)$ and $\psi(2S)-\eta_c(2S)$ splittings.  To reach
agreement with experiment it is shown to be small, $r_{ss} \leq 0.06$
fm. Our value $r_{ss}=0.057$  fm practically coincides with the number
used in Ref. \cite{10}.  However, in spite of this coincidence the
splitting $\Delta_1=\Upsilon (1S) -\eta_b(1S) =28 $ MeV in our
calculations appears to be two times smaller than that in Ref.
\cite{10}, where $\Delta_1 =60$ MeV.

From our point of view the use of the w.f. at the origin  $|\tilde
R_n(0)|^2_{\exp}$, extracted from  di-electron widths, is the most
promising one, because these w.f. take implicitly into account the
relativistic corrections as well as the influence of open channel(s),
in this way drastically simplifying the theoretical analysis. A
comparison of these w.f. with those calculated in different models puts
serious restrictions on the static potential used and also on
many-channel models.

The HF interaction is considered here in two cases. First one
corresponds to the standard perturbative (P) spin-spin interaction with
a $\delta$-function:
\be
 \hat V^{\rm P}_{ss} (r) =
 \ves_1\cdot\ves_2 \frac{32\pi}{9\omega_Q^2} \alpha_s (\tilde \mu)
 \left(1+\frac{\alpha_s}{\pi}\rho\right)\delta (\ver)
 \equiv\ves_1\cdot \ves_2 V_{\rm HF}(r),\label{1a}
\ee
which in one-loop approximation gives the following  HF splitting \cite{6}:
\be 
 \Delta^{\rm P}_{\rm HF} (nS) =
 \frac89 \frac{\alpha_s(\tilde \mu)}{\omega^2_Q} |R_n(0)|^2
 \left[1+\frac{\alpha_s(\tilde \mu)}{\pi}\rho \right],
\label{2}
\ee  
where $\rho=\frac{5}{12} \beta_0 -\frac83-\frac34 \ln 2$ and the second
term in brackets is small: $ \la 0.5\%$ in bottomonium $(n_f=5)$ and
$\la 3\% $  in charmonium $(n_f =4) $.

It is very probable that $\delta(\ver)$ may be considered as a limiting
case and the ``physical" spin-spin interaction is smeared with a still
unknown ``smearing" radius. For the Gaussian smearing function
\be
 \delta(\ver) \to \frac{4\beta^3}{\sqrt{\pi}}\int r^2 dr \exp (-\beta^2 r^2)
\label{4} 
\ee 
the splitting can be rewritten as
\be
 \Delta^{\rm P}_{\rm HF} (nS) =\frac89 \frac{\alpha_s(\tilde \mu)}
 {\omega^2_Q}\xi_n(\beta) |R_n(0)|^2 \left( 1+\frac{\alpha_s}{\pi}\rho\right), 
\label{5} 
\ee 
where by definition the ``smearing factor" $\xi_n (\beta)$ is
 \be \xi_n (\beta) =\frac{4}{\sqrt{\pi}}
 \frac{\beta^3}{|R_n(0)|^2} \int |R_n(r)|^2 \exp (-\beta^2r^2) r^2 dr.
\label{6}
\ee
The general  expression (\ref{5}) is evidently kept for  any other
smearing prescription which may differ from Eq.~(\ref{4}).

\section{Wave function at the origin}  
The w.f. at the origin is very sensitive to the form and parameters of
the gluon-exchange interaction and also to the value of the quark mass used.
Therefore we make the following remarks:
\begin{description}
\item [a.] ~~To minimize the uncertainties in the w.f. at the origin,
$R_n(0)$,  we shall use the w.f. extracted from the experimental data
on leptonic widths and denote them as $|\tilde R_n(0)|^2_{\rm exp}$. In
this way the relativistic corrections to the w.f. and the influence of
open channel(s) are implicitly taken into account.

\item [b.]
~~In Eqs.~(\ref{2}) and (\ref{5}) the constituent mass $\omega_q$
enters (this fact can  be rigorously deduced from relativistic
calculations \cite{14,15}):  $ \omega_q (nS) =\lan \sqrt{\vep^2 +
m^2_Q}\ran_{nS}$, where under the square-root the pole mass $m_Q\equiv
m_Q $ (pole) is present.  This mass is known with good accuracy  and we
take here $m_b({\rm pole}) =4.8 \pm 0.1$ GeV and $m_c ({\rm pole})
=1.42\pm 0.03$ GeV, which correspond to the well-established current masses
$\bar m_b (\bar m_b)=4.3(1)$ GeV, $\bar m_c(\bar m_c) =1.2(1)$ GeV
\cite{3}, while the constituent masses lie also in a rather narrow
range for all $nS$ states, both in charmonium and bottomonium:  $
\omega_b(nS)=5.05 \pm 0.15~{\rm GeV}$, $\omega_c(nS)=1.71 \pm 0.03~{\rm
GeV} (n\geq 2)$, and $\omega_c(1S) =1.62 \pm 0.04~{\rm GeV} (n=1)$.
Note, that just these mass  values are mostly used in nonrelativistic
calculations, thus implicitly taking into account relativistic
corrections.

\item [c.] ~~The leptonic width of the $n~^3S_1$ states in heavy
quarkonia are defined  by the Van Royen-Weisskopf formula with QCD
correction $\gamma_Q$,
\be
 \Gamma_{ee} (n~^3S_1)|_{\exp} =\frac{4e^2_Q\alpha^2}{M^2_n}
 |\tilde R_n(0)|^2_{\rm exp} \gamma_Q,  
\label{a8} 
\ee 
\end{description} 
where $e_Q=\frac13 \left( \frac23\right)$ for a $b(c)$ quark,
$\alpha=(137)^{-1}$, $M_n\equiv M(n~^3S_1)$, and $\gamma_Q(nS)
=1-\frac{16}{3\pi}\alpha_s(2m_Q),$  with  the renormalization scale
$\mu$ in  $\alpha_s$ equal to $2m_Q$~(pole), as in Refs.~\cite{9,10}
and also in $\eta_b \to \gamma\gamma$ decay \cite{16a}.  In some cases
$\mu=M_n$ is also taken, but with accuracy $\la 1\%$ both choices
coincide (here $2m_b=9.6$ GeV and $2m_c=2.9$ GeV are taken).

Since for $n_f=5$ the QCD constant $\Lambda^{(5)}_{ \overline{MS}}$ is
well known from  high energy experiments \cite{3}, the factor
$\gamma_b$ is also defined with a good accuracy. For  $\Lambda^{(5)}_{
\overline{MS}}(3-loop) =210(10)$ MeV, which corresponds to $\alpha_{
\overline{MS}}(M_Z)=0.1185$,  one has $ \gamma_{b}=\gamma_{bn}
=0.700(5) $ and $\alpha_s(2m_b) =0.177(3).  $ In charmonium $(n_f=4)$
for $\Lambda^{(4)}_{ \overline{MS}} =0.260(10)$ MeV the coupling $
\alpha_s (2m_c=2.9\, {\rm GeV})=0.237(5)$, $\gamma_c =0.60 (2)$.  Then
the w.f. at the origin, extracted from the di-electron width (\ref{a8}),
\be |\tilde R_n(0)|^2_{\rm exp} =\frac{M^2_n\Gamma_{ee}
(n~^3S_1)}{4e^2_Q \alpha^2\gamma_{Q}},\label{12}
\ee
implicitly takes into account the relativistic corrections as well as
the influence of open channels, which gives rise to smaller values for
$|R_n(0)|$ as well as for the HF splitting.  The extracted  values of
$|\tilde R_n(0)|^2_{\rm exp}$ in the $b\bar b$ and $c\bar c$ systems are
presented in Table~\ref{table1}.

\begin{table}
\caption{The w.f. $|\tilde R_n(0)|^2_{\rm exp}$ (in GeV$^3$) and the
leptonic widths $\Gamma_{ee} (\Upsilon(nS))$ and $\Gamma_{ee}
(\psi(nS))$ (in keV)$^{a,b)}$  ($\gamma_b=0.70$, $\gamma_c=0.60)$.
\label{table1}}
\begin{center}
\begin{tabular}{|c|c|c|c|c|}
\hline
&\multicolumn{2}{c|}{~}&\multicolumn{2}{c|}{~}\\
&\multicolumn{2}{c|}{$b\bar b$}&\multicolumn{2}{c|}{$c\bar c$}\\\cline{2-5}&&&&\\
 &~$\Gamma_{ee} (nS)_{\rm exp}$~&~$|\tilde R_n(0)|^2_{\rm exp}~$&
~$\Gamma_{ee} (nS)_{\rm exp}$ ~&~$|\tilde R_n(0)|^2_{\rm exp}$\\
\hline
&&&&\\
&$1.314(29)^{a)}$&7.094(16)& 5.40(22)& 0.911(37)\\
$1S$&&&&\\
&1.336(28)$^{b)}$&7.213(15)&5.68(24)&0.959(40\\
\hline
&& &&\\
&0.576(24)$^{a)}$& 3.49(15)&2.12(12)&0.51(3)\\\
$2S$&&&&\\
&0.616(19)$^{b)}$& 3.73(12)&2.54(14)&0.61(3)\\
\hline
&&&&\\
$3S$&0.413(10)$^{b)}$& 2.67(7)&0.75(1)&0.22(1)
\\&&&0.89(8)&0.26(2)\\
\hline
&&&&\\
$4S$&0.25(3)$^{a)}$&1.69(20)&0.47(15)&0.16(5)\\
&&&0.71(10)&0.24(4)\\
\hline
&&&&\\
$5S$& 0.31(7)$^{a)}$ &2.21(49)&&\\
&&&&\\
\hline 
&&&&\\
$6S$&0.13(3)$^{a)}$&0.95(22)&&\\ 
&&&&\\
\hline
\end{tabular}
\end{center}
$^{a)}$ The upper values of the leptonic widths in bottomonium  are taken
from PDG~\cite{3} and the lower values of $\Gamma_{ee} (nS)$ are taken
from the CLEO data~\cite{12}.

$^{b)}$ The upper entries in charmonium are taken from [3] and the
lower ones from \cite{13}.
\end{table}

The extracted  $|\tilde R_n(0)|^2_{\rm exp}$ can be compared to the
predicted values, which chiefly depend on the  strong coupling used in
the gluon-exchange term. In particular, if the asymptotic-freedom
behavior of $\alpha_{\rm static} (r)$ is neglected, then theoretical
numbers can be $2 - 1.5$ times larger than $|\tilde R_n(0)|^2_{\rm
exp}$, even for the $\Upsilon (nS)\,(n=1,2,3)$ states, which lie far
below the $B\bar B$ threshold \cite{8}.

Here, as well as in our analysis of the spectra and fine structure
splittings in heavy quarkonia \cite{7,14,15}, we  use the static
potential $V_B(r)$ in which the strong coupling $\alpha_B(r)$ is
defined as in background perturbation theory:  
\be
 V_B(r) =\sigma r -\frac43 \frac{\alpha_B(r)}{r},\quad
\alpha_B(r) =\frac{8}{\beta_0}\int dq\frac{\sin qr}{q}
 \frac{1}{t_B(q)}
\left[ 1-\frac{\beta_1}{\beta_0^2} \frac{\ln t_B}{t_B}\right],
\label{A.2}
\ee 
%
%with the coupling $\alpha_B(r) =\frac{8}{\beta_0}\int dq\frac{\sin qr}{q}
% \frac{1}{t_B(q)}
%\left[ 1-\frac{\beta_1}{\beta_0^2} \frac{\ln t_B}{t_B}\right]$, 
where $t_B(q) =\ln \frac{q^2+M^2_B}{\Lambda^2_B(n_f)}$. Here $M_B =0.95
(5)$ GeV is the background mass,  $\Lambda_B (n_f)$ is  expressed
through $\Lambda_{\overline{MS} }(n_f)$ and in 2-loop approximation
$\Lambda_B (n_f =4) =360(10)$ MeV  and $\Lambda_B(n_f=5) =335(5)$ MeV
\cite{14}; the  string tension $\sigma=0.18$ GeV$^2$.  Our calculations
show that in bottomonium (in single-channel approximation) the
potential $V_B(r)$ gives values of $|R_n(0)|^2_{\rm theory}$  very
close to the values $|\tilde R_n(0)|^2_{\rm exp}$.  For illustration
in Table~\ref{table2} the ratios
\be
 S_n =\frac{|\tilde R_n(0)|^2_{\rm exp}}{|\tilde R_n(0)|_{\rm theory}^2}
\label{13}
\ee 
are given for all known $nS$ levels in charmonium and bottomonium.

\begin{table}
\caption{The factor $S_n$ (9) for the potential $V_B(r)$ (8) in
charmonium and bottomonium.\label{table2} }
\begin{center}
\begin{tabular}{|c|c|c|c|c|c|c|}
\hline
&&&&&&\\
&~$1S$~&~$2S$~&~$3S$~&~$4S$~&~$5S$~&~$6S$~\\ 
&&&&&& \\
\hline
&&&&&&\\
$b\bar b$ &1.08(4) &1.02(4)&1.02(4)&$ 0.72 (9)$&1.03(22)&0.47(10)\\
&&&&&&\\
\hline
&&&&&&\\
$c\bar c$&1.01(4)&0.82(5)&$0.41(2)$&$0.32(10)$&&\\
&&&&&&\\
\hline
\end{tabular}
\end{center}
\end{table}

As seen from Table~\ref{table2}, using potential $V_B(r)$ the influence
of open channels in bottomonium appears to be important only for the
$4S$ and $6S$ levels, while for the other states single-channel
calculations are in good agreement with experiment. This is not so for
many other potentials \cite{8} and it means that any conclusions about
the role of open  channels cannot be separated from the $Q\bar Q$
interaction used in a given theoretical approach.

In charmonium the effect from open channels is much stronger and
already reaches  $\sim  60\%$ for the $3S$ and the $4S$ states 
($S_n \approx 0.4$) and about 20\% for the $\psi (2S)$ meson.

\section{Hyperfine splitting} 
Now we discuss the HF splitting for both bottomonium and charmonium.

\subsection{Bottomonium}
In bottomonium  the HF splittings are considered in two cases:
\begin{description} 

\item[A.] No smearing  effect, i.e., in Eq.~(\ref{5}) the smearing
parameter $\xi_{bn} =1.0\;(\forall n)$.

\item[B.] The smearing parameter $\xi_{bn}$ (\ref{6}) is calculated
with $\beta=\sqrt{12}$ GeV, or a smearing radius $r_{ss}=\beta^{-1}
=0.057$ fm.  

\end{description} 
Unfortunately, at present there is no a precise prescription how to
choose the renormalization scale in the HF splitting (2):  in
$\alpha_{\overline{MS}}(\tilde \mu)$ the scale $\tilde \mu=m_b({\rm
pole}) \approx 4.80 \pm 0.01$ GeV is often used.  With $\Lambda_{
\overline{MS}}(n_f=5)=210 (10)$ MeV (just the same as in our
calculations of $\gamma_b$ (6)) one finds
\be
 \alpha_s(b\bar b,\tilde \mu) =\alpha_{\overline{MS}} (4.8 ~{\rm GeV})
 =0.21(1).  
\label{14} 
\ee 
With this $\alpha_s(\tilde \mu)$ and $|\tilde R_n(0)|^2_{\rm exp}$ from
Table~\ref{table1}, one obtains the HF splittings in bottomonium
presented in Table~\ref{table3}, second column. (The numbers in Table 3
contain experimental errors coming from $\Gamma_{ee} (nS)$ \cite{3}
(first number) and theoretical errors (second number)).  For  the
$\Upsilon (nS)$ states $ (n=1,2,3)$  the calculated  HF splittings
($\xi_n=1.0$) appear to be  very close to the splittings from Refs.
\cite{9}.

If smearing of the HF interaction (3) is taken into account (the
smearing radius, $r_{ss}=\beta^{-1}=0.057$ fm  for $\beta
 =\sqrt{12}$ GeV, is taken to  fit the experimental values of the
$J/\psi-\eta_c(1S)$ and  $\psi(2S)-\eta_c(2S)$ splittings), then even
for such a small radius  $\Delta_{\rm HF}(nS)$ turn out to be 50\%
$(n=1,\dots,4),$ 60\% $(n=5,6)$ smaller as compared to the ``nonsmearing"
case. In particular, the $\Upsilon $(1S)$ -\eta_b (1S)$ splitting turns
out to be  28 MeV instead of 51(4) MeV for  $\xi_{bn} =1.0$. For higher
excitations very small splittings, $\Delta_{\rm HF} \approx 6$ MeV and
3 MeV for the $5S$ and $6S$ states, are obtained, see
Table~\ref{table3}.

\begin{table}
\caption{$\Delta^{\rm P}_{\rm HF}(nS)$ (in MeV) in bottomonium for
$\alpha_{\overline{MS}}{(\tilde \mu)}=0.21$, $\omega_b=5.10$ GeV and
$|\tilde R_n (0)|^2$ from Table~\ref{table2}.\label{table3}}
\begin{center}
\begin{tabular}{|c|c|c|}
\hline
&&\\
 &~$\xi_b=1.0~$~&~$\xi_{bn}$ for the smeared HF interaction\\
&(no smearing)& with $\beta=\sqrt{12}$ GeV,\\
&& $r_{ss} = 0.057$ fm\\
&&\\
\hline
&&\\
$1S$&51(4) (4)& 28(2) (3)\\
&&\\
\hline 
&&\\
$2S$&25(3) (2) &12(2) (1)\\
&&\\
\hline
&&\\
$3S$&22(5)(2)& 10(2) (1)\\
&&\\
\hline
&&\\
$4S$&12(3) (1)&5.1(2) (1)\\
&&\\
\hline
&&\\
$5S$& 16(2) (1) &6.4(1) (1)\\
&&\\
\hline 
&&\\
$6S$&7(2) (1)&2.7(1) (1)\\ 
&&\\
\hline
\end{tabular}
\end{center}
\end{table}
Note that our value of $r_{ss}=0.057$ fm is very close to that from
Ref.~\cite{10} where $r_{ss}=0.060$ fm is taken.  However, in spite of
this coincidence our numbers are about two times smaller than in
\cite{10}, where  $\Upsilon(1S)-\eta_b(1S)$ = 60 MeV is obtained. For
the $2S$ state our value of the splitting is 12 MeV, still smaller than
20 MeV in \cite{10}. From this analysis it is clear that the
observation of an $\eta_b(nS)$ meson could clarify the role of smearing
in the spin-spin interaction between a heavy quark and antiquark.

\subsection{Charmonium}
Also in charmonium the splitting (\ref{5}) in fact depends on the
product $\alpha_s(\tilde \mu)\cdot \xi_n$, therefore it is convenient
to discuss an \underline{effective} HF coupling: $\alpha_{\rm HF} (nS)
=\alpha_s (\tilde \mu_n) \xi_{cn}$, which is the only unknown factor.
(The masses $\omega_c(nS)$ may be specified for different $nS$ states
[14,15].)

As discussed in \cite{7}, the experimental splittings $J/\psi -\eta_c $
(1S) and $\psi$ (2S) $-\eta_c$(2S) can be fitted if different values of
$\alpha_{\rm HF}$ for the 1S and 2S states are taken namely,
$\alpha_{\rm HF} $(1S) $\approx 0.36$ and $\alpha_{\rm HF} $(2S) $\approx
0.30.$ Such a choice implies two possibilities. The first one, case A, is
\begin{equation}
 A.\; \alpha_s(\mu_1) =0.36,~ \alpha_s(\mu_2) =0.30,~
 ~\alpha_s(\mu_3) =\alpha_s(\mu_4)\leq 0.30,~ \xi_{cn} =1.0 ~(\forall n),
\label{18}
\end{equation}
i.e., the renormalization scale is supposed to grow for larger
excitations. In particular,  for $\Lambda^{(4)}_{\overline{MS} }(2-{\rm
loop})=270$ MeV one finds $\mu_1=1.25$ GeV $\approx\bar m_c(\bar m_c)$
while the scale $\mu_2=1.60$ GeV is essentially larger. For this
choice of $\alpha_{\rm HF}$ the perturbative HF splittings are given in
Table~\ref{table4}, second column.

\begin{table}
\caption{The splittings $\Delta^{\rm P}_{\rm HF}(nS)$ and
$\Delta^{\rm NP}_{\rm HF} (nS) $ (in MeV) in charmonium$^{a)}$ \label{table4}}
\begin{center}
\begin{tabular}{|c|c|c|c|}
\hline
&&&\\
&~$\Delta^{\rm P}_{\rm HF}(nS)$~&~$\Delta^{\rm P}_{\rm HF}(nS)$~&\\
&(no smearing: $\xi_c=1.0)$&$r_{ss} =0.29$ GeV$^{-1}$ & 
$\Delta^{\rm NP}_{\rm HF}(nS)^{b)}$\\
&$\alpha_s(\mu_1)=0.36; \alpha_s(\mu_n) =0.30$ & $\alpha_s(\mu_n)=0.36$ &
$G_2=0.043$\\
&$(n=2,3,4)$&$(n=1,\dots,4)$&GeV$^4$\\
&&&\\
\hline
&&&\\
$1S$&117(5)& 102(6)$^{a)}$&$9\pm 2$\\&&108(7)$^{c)}$& \\
&&& \\
\hline
&&&\\
experiment& 117(2)& 117(2)&\\ $J/\psi-\eta_c$ (1S)&&&\\
&&&\\
\hline
&&&\\
$2S$&51(5)&46(3)&$3.5\pm 1.5$\\
&61(5)$^{c)}$&55(4)&\\
&&&\\
\hline
&&&\\
experiment &48(4)&48(4)&\\   $\psi$ (2S) $-\eta_c$ (2S) &&&\\
&&&\\
\hline
&&&\\
$3S$&21(2)& 16(2)&$2\pm 1$\\
&&&\\
\hline
 &&&\\
$4S$&15(4)&12(4)&$1.5\pm 0.5$\\
&&&\\
\hline
\end{tabular}
\end{center}
$^{a)}$ The w.f. $|\tilde R_n(0)|^2$, taken  from Table 1,
correspond  to  $\Gamma_{ee}(nS)$ from PDG\cite{3}.\\
$^{b)}$ The NP splittings are calculated in [17].\\
$^{c)}$ Here $|\tilde R_1(0)|^2_{\exp}=0.959 $ GeV$^3$ and $|\tilde
R_2(0)|^2 =0.61$GeV$^3$ from the CLEO data \cite{13} are taken.
\end{table}

Besides, we have also calculated the contributions coming from the NP
spin-spin interaction.  In bottomonium their values are small,
$\Delta_{\rm HF}^{\rm NP} (nS) < 1$ MeV, and can be neglected. In
charmonium, as well as in light mesons, the situation is different,
e.g. due to the NP spin-spin interaction in the $1P~c\bar c$ state a
cancellation of perturbative and NP terms takes place \cite{17}. As a
result, the mass difference $M_{\rm cog} (\chi_{cJ}) -M(h_c) =(+1\pm
1)$ MeV turns out to be close to zero or even positive, in accord with
experiment \cite{18}.  The values of $\Delta_{\rm HF}^{\rm NP}(nS)$ are
given in Table~\ref{table4}, fourth column.

Thus  one can conclude that in case $A$  with different renormalization
scales $\mu_n$, the splittings $J/\psi-\eta_c (1S)$ and $\psi (2S) -
\eta_c (2S)$ can be obtained easily in agreement with experiment.

If the  renormalization scales $\mu_n$  are  supposed to be  (almost)
equal for all $nS$ states:
\begin{equation}
 B. \quad \alpha_s(\mu_n\approx\bar m_c= 1.25\;{\rm  GeV}) =0.36 ,
\end{equation}
then to explain the relatively small $\psi (2S) - \eta_c (2S)$
splitting, a smearing effect needs to be introduced. Then for the
potential used, the values $\xi_n (c \bar c)= 0.84,\, 0.80,\, 0.78$, and $0.76$
for the $1S$, $2S$, $3S$, and $4S$ states, respectively, are calculated.
In this case the $\Delta_{\rm HF}^{\rm P} (c\bar c, nS)$ are also given
in Table~\ref{table4}. For the higher $3S$ ($4S$) levels our predicted
numbers are about 21(15) MeV (no smearing) and 16(12) MeV (with
smearing), i.e., the difference between cases $A$ and $B$ is only $\sim
20\% $. Notice that in case $B$  the NP contribution improves the
agreement with experiment for $J/\psi-\eta_c(1S)$. As a whole, in
charmonium the smearing effect appears to be less prominent than in
bottomonium.

\section{Conclusions}
Thus we come to the following conclusions:

\begin{enumerate}
\item In bottomonium $\Delta_{\rm HF}^{\rm P}(nS)$ appears to be  very
sensitive to the smearing of the spin-spin interaction.  Due to this effect
the splitting decreases from 51 MeV to 28 MeV for the $1S$ state and
from 25 MeV to 12 MeV for the $2S$ state; very small values are
obtained for higher states.

\item  In charmonium there are two possibilities to describe
$\Delta_{\rm HF} (1S)$ and $\Delta_{\rm HF} (2S)$, which are  known
from experiment. First one refers to a different choice of the
renormalization scale: $\mu_1$ =1.25 GeV and $\mu_2\approx 1.60$ GeV for
the $1S$ and $2S$ states, if the smearing effect is absent. The second
possibility implies equal renormalization scales $\mu_n (n=1,\dots, 4)$ 
for all $nS$ states. Then to explain the $\psi(2S) -\eta_c(2S)$
splitting the smearing of the spin-spin interaction needs be taken
into account. We also  expect that  for the $1S$ level a small
contribution ($\sim $ 8 MeV) comes from the NP spin-spin interaction.

\item The $\psi(3S) - \eta_c(3S)$ splitting in single-channel
approximation is predicted to be around 16(2) MeV, without and 12(4)
MeV with smearing effect.  
 
\end{enumerate} 
In order to understand the true role of the smearing effect in the
spin-spin interaction the observation of an $\eta_b(nS)$ is crucially
important.

\end{document}